\documentclass[a4paper]{article}

\usepackage{graphicx}
\usepackage{xspace}
\graphicspath{ {./img/} }

\usepackage[utf8]{inputenc}
\usepackage{amsmath}
\usepackage{amssymb}
\usepackage{color}

\usepackage{tabu}
\usepackage{multirow}
\usepackage{booktabs}
\usepackage{sidecap}
\usepackage[binary-units]{siunitx}

\usepackage{subcaption}
\usepackage{hyperref}

\usepackage[thref,amsmath]{ntheorem}

\newtheorem{lemma}{Lemma}

\newtheorem{definition}{Definition}

\newtheorem{problem}{Problem}

\theoremstyle{nonumberplain}
\theoremheaderfont{\itshape}
\theorembodyfont{\normalfont}
\theoremseparator{.}
\theoremsymbol{}
\newtheorem{proof}{Proof}

\providecommand{\keywords}[1]{\textbf{\textit{Keywords ---}} #1}

\usepackage{authblk}

\newbox{\myorcidaffilbox}
\sbox{\myorcidaffilbox}{\large\includegraphics[height=1.7ex]{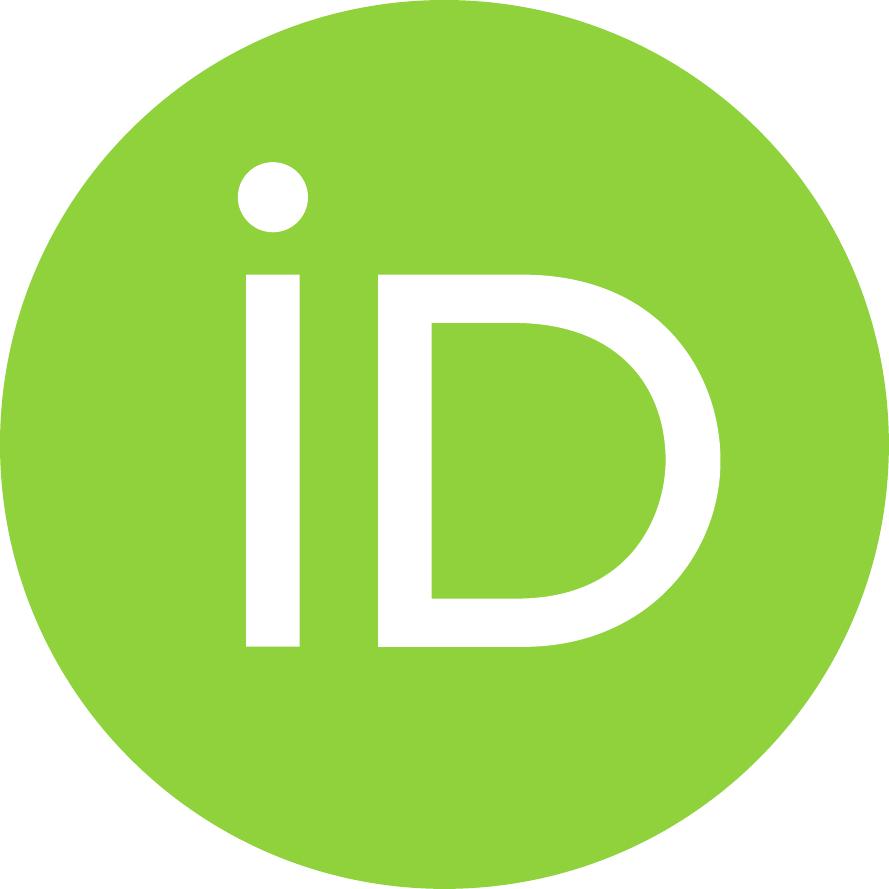}}
\newcommand{\orcid}[1]{%
	\href{https://orcid.org/#1}{\usebox{\myorcidaffilbox}}}

\title{Tailoring r-index for metagenomics}%

\author[1]{Dustin Cobas\orcid{0000-0001-6081-694X}}
\author[2]{Veli Mäkinen\orcid{0000-0003-4454-1493}}
\author[3]{Massimiliano Rossi\orcid{0000-0002-3012-1394}}

\affil[1]{CeBiB --- Center for Biotechnology and Bioengineering, Department of Computer Science, University of Chile, Santiago, Chile}
\affil[2]{Department of Computer Science, University of Helsinki, Helsinki, Finland}
\affil[3]{Department of Computer and Information Science and Engineering, University of Florida, Gainesville, USA}
{
    \makeatletter
	\renewcommand\AB@affilsepx{, \protect\Affilfont}
	\makeatother
	\affil[1]{dcobas@dcc.uchile.cl}
    \affil[2]{veli.makinen@helsinki.fi}
    \affil[3]{rossi.m@ufl.edu}
}

\date{December 2019}

\begin{document}

\def\SA{\textsf{SA}\xspace}
\def\CSA{\textsf{CSA}\xspace}
\def\RLCSA{\textsf{RLCSA}\xspace}
\def\ISA{\textsf{ISA}\xspace}
\def\LCP{\textsf{LCP}\xspace}
\def\ILCP{\textsf{ILCP}\xspace}
\def\VILCP{\textsf{VILCP}\xspace}
\def\LILCP{\textsf{LILCP}\xspace}
\def\BWT{\textsf{BWT}\xspace}
\def\ri{\emph{r}-index\xspace}
\def\DA{\textsf{DA}\xspace}
\def\bL{\textsf{L}\xspace}
\def\bR{\textsf{R}\xspace}
\def\bB{\textsf{B}\xspace}

\def\mSA{\ensuremath{\mathsf{SA}}}
\def\mCSA{\ensuremath{\mathsf{CSA}}}
\def\mRLCSA{\ensuremath{\mathsf{RLCSA}}}
\def\mISA{\ensuremath{\mathsf{ISA}}}
\def\mLCP{\ensuremath{\mathsf{LCP}}}
\def\mILCP{\ensuremath{\mathsf{ILCP}}}
\def\mVILCP{\ensuremath{\mathsf{VILCP}}}
\def\mLILCP{\ensuremath{\mathsf{LILCP}}}
\def\mBWT{\ensuremath{\mathsf{BWT}}}
\def\mC{\ensuremath{\mathsf{C}}}
\def\mB{\ensuremath{\mathsf{B}}}
\def\mL{\ensuremath{\mathsf{L}}}
\def\mR{\ensuremath{\mathsf{R}}}
\def\mAL{\ensuremath{\mathsf{AL}}}
\def\mAR{\ensuremath{\mathsf{AR}}}
\def\mDA{\ensuremath{\mathsf{DA}}}
\def\mDSA{\ensuremath{\mathsf{DSA}}}
\def\mDISA{\ensuremath{\mathsf{DISA}}}

\def\T{{\cal T}}
\def\S{{\cal S}}
\def\C{{\cal C}}
\def\D{{\cal D}}
\def\Oh{{\cal O}}

\def\ndoc{\texttt{ndoc}}
\def\rank{\texttt{rank}}
\def\select{\texttt{select}}

\newcommand\vsTColl{\mathit{n}}
\newcommand\vnDocs{\mathit{d}}
\newcommand\vsPat{\mathit{m}}
\newcommand\occ{\mathit{occ}}

\newcommand\idxName{\textsf}

\maketitle              %
\begin{abstract}
A basic problem in \emph{metagenomics} is to assign a sequenced read to the correct species in the reference collection. In typical applications in genomic epidemiology and viral metagenomics the reference collection consists of set of species with each species represented by its highly similar strains. It has been recently shown that accurate read assignment can be achieved with $k$-mer hashing-based \emph{pseudoalignment}: A read is assigned to species A if each of its $k$-mer hits to reference collection is located only on strains of A. We study the underlying primitives required in pseudoalignment and related tasks. We propose three space-efficient solutions building upon the \emph{document listing with frequencies} problem. All the solutions use an $r$\emph{-index} (Gagie \emph{et al.}, SODA 2018) as an underlying index structure for the text obtained as concatenation of the set of species, as well as for each species. Given $t$ species whose concatenation length is $n$, and whose Burrows-Wheeler transform contains $r$ runs, our first solution, based on a grammar-compressed document array with precomputed queries at non terminal symbols, reports the frequencies for the $\ndoc$ distinct documents in which the pattern of length $m$  occurs in $\Oh(m + \log(n)\ndoc) $ time. Our second solution is also based on a grammar-compressed document array, but enhanced with bitvectors and reports the frequencies in $\Oh(m + ((t/w)\log n + \log(n/r))\ndoc)$ time, over a machine with wordsize $w$. Our third solution, based on the interleaved LCP array, answers the same query in $\Oh(m + \log(n/r)\ndoc)$. 
We implemented our solutions and tested them on real-world and synthetic datasets. The results show that all the solutions are fast on highly-repetitive data, and  the size overhead introduced by the indexes are comparable with the size of the $r$-index.

\keywords{Metagenomics,  \and r-index, \and document listing.}
\end{abstract}

\section{Introduction}
Metagenomics is the study of genomic material recovered directly from environmental samples. Thus, conversely to genomic samples, metagenomic samples consist of genome sequences of a community of organisms sharing the same environment, highlighting the microbial diversity in the environmental samples. The samples of genome sequences are collected using shotgun sequencing. This creates a mixture of genome fragments from all organisms in the environment. One important step in metagenomics is to assign each fragment to its owner, allowing to identify and quantify species. This step is called read assignment \cite{huson2007megan}, and it is the basic step in most metagenomic analysis workflows such as in genomic epidemiology~\cite{maklin2020genomic}, and viral epidemiology~\cite{carroll2018global}. %

Read assigners were first implemented using computational expensive read aligners~\cite{huson2007megan,xia2011accurate,lindner2013metagenomic}. In~\cite{wood2014kraken} the authors showed that similar results are achieved replacing the read aligners with the less computational expensive $k$-mer hashing methods. Read assigners based on $k$-mer set indexing are referred to as \emph{pseudoaligners}. Efficient indexing of $k$-mer sets has been deeply investigated and we refer the reader to the survey~\cite{marchet2019data} for further reading. Pseudoaligners such as Kallisto~\cite{bray2016near}, MetaKallisto~\cite{Schetal17}, and Themisto~\cite{maklin2020genomic} are based on the following pseudoalignment criterion. Given a set of references $T_1, \ldots, T_t$ (representing $t$ distinct species), and read $P$, the read $P$ is pseudoaligned with $T_i$ if there exists a $k$-mer of $P$ that occurs in $T_i$ and for all other $k$-mers $u$ of $P$, either $u$ occurs in $T_i$ or $u$ does not occur in $T_1, \ldots, T_t$. %
This approach and its solutions using {\em colored de Bruijn graphs}~ \cite{bray2016near,Schetal17,maklin2020genomic} are motivated by the fact that the species are usually quite dissimilar, but the strains inside the species are highly similar.

In this paper, we study some basic primitives that are required in different variations of the pseudoalignment criteria. We argue that the specific criterion given above is just one example of a family of criteria, and it is important to study the general framework rather than tailoring the methods to a very narrow setting. 
Towards this goal of obtaining general results, instead of studying directly $k$-mers of a pattern, we focus here on searching the complete pattern. We continue the discussion in Sect.~\ref{sect:discussion} on how to integrate the results with $k$-mer based criteria.

We modelled this read assignment problem as a {\em document listing with frequencies} problem, where the set of species is a collection and each species is a document formed by the concatenation of its strains. Given a pattern $P$ we want to report all documents where $P$ occurs, and their frequencies. This problem was first introduced in~\cite{DBLP:conf/cpm/ValimakiM07} and further refined in~\cite{DBLP:journals/jda/BelazzouguiNV13} and~\cite{DBLP:journals/tcs/GagieNP12} (details in Sect.~\ref{sect:related work}). We propose three solutions. All solutions use an $r$-index~\cite{Travis18} as text index for the concatenation of all documents.
 The first solution is an extension to frequencies of the solution proposed in~\cite{cobas2019fast} in which a grammar-compressed document array is used, and for each non terminal node, precomputed answers are stored. The second and the third solution are based on the {\em term frequency} approach presented in~\cite{sadakane2007succinct} which uses an additional index for all documents. The key idea is to find the leftmost and rightmost occurrence of the pattern $P$ in the index of each document, by searching the pattern in the index of the concatenation of all documents. To do this, the second solution uses the grammar-compressed document array of~\cite{cobas2019fast} enhanced with bitvectors at non terminal nodes marking which descendant contains the leftmost and rightmost occurrence of the pattern in each document. The third solution relies on a modified version of the interleaved longest common prefix array~\cite{gagie2017document}. We implemented our solutions and we tested them using real-world and synthetic datasets. 

\section{Basics}
A string $S$ is a sequence of characters over an alphabet $\Sigma$ of size $\sigma = |\Sigma|$. 
A document $T$ is a string terminated by a special symbol $ \$\notin \Sigma$ that is lexicographically smaller than all characters in $\Sigma$. A collection $D = \{T_1,T_2,\ldots,T_t\}$ is a set of $t$ documents, which is usually represented as the concatenation of its documents, i.e. $\D = T_1 T_2 \cdots T_t$.
When it is clear from the context, we will refer to $T_i$ as document $i$. 
Given a string $S[1..n]$, let $\rank_c(S,i)$ be the number of occurrences of symbol $c$ in $S[1..i]$, and let $\select_c(S,j)$ be the position of the $j$-th symbol $c$ in $S[1..n]$.
When string $S$ is from alphabet $\{0,1\}$, we call it a bitvector. For bitvector $S$ 
it holds $\rank_0(S,i)=i-\rank_1(S,i)$. 

Given a string $S$ over an alphabet $\sigma$, the {\em suffix array}~\cite{DBLP:journals/siamcomp/ManberM93} $\mSA[1..n]$ of $S$ is an array of integers providing the starting position of the suffixes of $S$ sorted in lexicographic order. The {\em inverse suffix array} $\mISA[1..n]$ of $S$ is an array of integers that, for each suffix of $S$, provides the position of the suffix in the suffix array. In particular we have that for all $1 \leq i \leq n$, $\mSA[\mISA[i]] = i$.

A {\em compressed suffix array}~\cite{DBLP:journals/csur/NavarroM07} $\mCSA[1..n]$ are space-efficient representations of the suffix array whose size $|\mCSA|$ in bits is usually bounded by $\Oh(n\log\sigma) $. We denote by $t_{search}(m)$ the time to find the interval of the suffix array corresponding to all occurrences of $P[1..m]$, while by $t_{lookup}(n)$ the time necessary to access any value $\mSA[i]$.

The \ri~\cite{Travis18} is a compressed text index whose main components are a run-length encoded {\em Burrows-Wheeler} transform (\BWT)~\cite{BW94} and the sample of the suffix array at the beginning and at the end of each run of the \BWT. The $r$-index can be computed in $\Oh(n)$ time and occupies $\Oh(r\log(n/r))$ space. We can find all occurrences of a given pattern $P[1..m]$ in the text $S[1..n]$ in time $\Oh(m + occ)$ time.
The $r$-index supports \SA and \ISA queries in $\Oh(\log(n/r))$ time and $\Oh(r\log(n/r))$ space.

Given a collection $D=\{T_1, \ldots, T_t\}$ of $t$ documents and its concatenation ${\cal D} = T_1 T_2 \cdots T_t$, the {\em document array}~\cite{muthukrishnan2002efficient} $\mDA[1..n]$ stores in each position $i$ the index of the document of which the suffix $\mSA[i]$ belongs to. 

Given a text $T[1..n]$, the {\em longest common prefix} array $\mLCP_T[1..n]$ stores in each position $2 \leq i \leq n$ the length of the longest common prefix between the two strings $T[\mSA[i-1]..n]$ and $T[\mSA[i]..n]$.

Given a collection $D = \{T_1, \ldots, T_t\}$ whose concatenation is $\D[1..n]$, the interleaved longest-common-prefix array $\mILCP[1..n]$ is defined in~\cite{gagie2017document} as the interleaving of the \LCP arrays of the documents $T_1, \ldots, T_t$ in the order they appear in the suffix array of $\D$, i.e., if $\mSA[i]$ is the lexicographically $j$-th suffix of the $k$-th document, $\mILCP[i] = \mLCP_k[j]$.
Let the \ILCP array be run-length encoded in $\rho$ runs. Then, it can be represented using two arrays: $\mLILCP[1..\rho]$ the prefix sums of the lengths of the $\rho$ runs; $\mVILCP[1..\rho]$ contains the values of these runs. Furthermore, the \LILCP array can be replaced by a sparse bitvector $\mL[1..n]$ such that $\mLILCP[i] = \select_1(\mL,i)$. 

Given a string $S[1..n]$, a {\em straight line grammar} for $S$ is a context-free grammar ${\cal G}$ that uniquely generates the string $S$. We denote by $\T$ the parse tree of $S$. Given a node $t \in \T$, $t$ is a {\em terminal} node if $t$ has no children, $t$ is a {\em non terminal} node otherwise. 
Each node $t \in \T$ uniquely identify an interval of $S$ denoted by $S[\ell_t..r_t]$. For the ease of explanation we say that a character $c$ occurs in $t$ by meaning that the character $c$ occurs in $S[\ell_t..r_t]$. The parse tree $\T$ is {\em binary} if its maximum arity is $2$, and $\T$ is {\em balanced} if every substring is covered by $\Oh(\log n)$ maximal nodes. 
Computing the smallest grammar is an NP-hard problem~\cite{lehman2002approximation}, but various $\Oh(\log(n/{\cal G}^\ast))$-approximation exists. We consider those that are binary and balanced~\cite{DBLP:journals/tcs/Rytter03,DBLP:journals/tit/CharikarLLPPSS05,DBLP:journals/tcs/Jez16}.

\section{Related Work}\label{sect:related work}

In this section we define three problems and report solutions and techniques from the literature that are used in our approach. For a complete overview we refer the reader to the survey~\cite{navarro2014spaces}. 
\begin{problem}[Document listing]\label{problem:listing}
Given a collection $D = \{T_1,T_2,\ldots,T_t\}$, and a pattern $P$, return the set of documents $L \subseteq D$ where $P$ occurs.
\end{problem}

Muthukrishnan~\cite{muthukrishnan2002efficient} proposed the first solution to Problem~\ref{problem:listing} in optimal time and linear space. He defined the {\em document array} $\mDA$ and used a suffix tree~\cite{DBLP:conf/focs/Weiner73} to find all occurrences of the pattern $P$ represented as an interval $[s_p..e_p]$. Then, he proposed a recursive algorithm to find all distinct documents $\ndoc$ in $\DA[s_p..e_p]$ in optimal time $\Oh(\ndoc)$. An extended description can be found in Appendix~\ref{app:Muthukrishnan}.

Sadakane~\cite{sadakane2007succinct} replaced the suffix tree with a compressed suffix array $\mCSA$ and the document array with a bitvector marking the starting position of each document in text order. He also replace the data structures to find all distinct documents $\ndoc$ in $\DA[s_p..e_p]$ with a succinct version using only $\Oh(n)$ bits. With this solution, Problem~\ref{problem:listing} can be solved in $\Oh(t_{search}(m) + \ndoc t_{lookup}(n))$ using a data structures of $|\mCSA| + \Oh(n)$ bits. An extended description can be found in Appendix~\ref{app:Sadakane}.

Gagie {\em et al.}~\cite{gagie2017document} introduced the \ILCP array whose property stated in Lemma~\ref{lemma:ilcp} allows to apply almost verbatim the technique used by Sadakane to find distinct elements in $\mDA[s_p..e_p]$. The solution uses a run-length compressed suffix array \RLCSA~\cite{MakinenNSV10:SRH} which allows to answer the queries of Problem~\ref{problem:listing} in $\Oh(t_{search}(m) + \ndoc t_{lookup}(n))$ time. An extended description can be found in Appendix~\ref{app:Gagie et al}.

Claude and Munro~\cite{claude2013document} proposed the first grammar-based document listing, later improved by Navarro in~\cite{navarro2019document}. Cobas and Navarro~\cite{cobas2019fast}, later proposed a practical variant in which they store the {\em document array} as a binary balanced straight line grammar. Then, they precompute and store the answers for all non terminal nodes of the grammar. The queries are answered by using a \CSA to find the interval $\mDA[s_p..e_p]$ and merging the precomputed answers for the $\Oh(\log n)$ non terminal symbols covering $\mDA[s_p..e_p]$. This leads to a solution that solves Problem~\ref{problem:listing} in $\Oh(t_{search}(m) + \ndoc\log n)$ time.

\begin{problem}[Term frequency]\label{problem:freq}
Given $D = \{T_1,T_2,\ldots,T_t\}$, and a pattern $P$, for each document $T \in D$ return the number of occurrences of $P$ in $T$.
\end{problem}

Sadakane~\cite{sadakane2007succinct}, addressed also the {\em term frequency} problem. The solution to Problem~\ref{problem:listing} is enhanced building a compressed suffix array $\mCSA$ for each document. Given the interval $[s_p..e_p]$ of all occurrences of the pattern $P$, he uses the data structure to find the distinct documents in $\DA[s_p..e_p]$ to find the leftmost occurrences of these documents. In a similar way he locate also the rightmost occurrences. Those positions are then mapped into an interval in the $\mCSA$ of the document. The sizes of these intervals represent the frequencies of the documents. This approach solves Problem~\ref{problem:freq} in $\Oh(t_{search}(m) + \ndoc\cdot t_{lookup}(n))$ time.

\begin{problem}[Document listing with frequencies]\label{problem:listing+freq}
Given $D = \{T_1,T_2,\ldots,T_t\}$, and a pattern $P$, return the set of documents where $P$ occurs and their frequencies.
\end{problem}

V{\"a}lim{\"a}ki and M{\"a}kinen~\cite{DBLP:conf/cpm/ValimakiM07} first proposed Problem~\ref{problem:listing+freq} and showed that the document listing problem can be solved using a rank and select data structure on the document array, to simulate Muthukrishnan's~\cite{muthukrishnan2002efficient} solution. %
In addition, after locating the interval $\mSA[s_p..e_p]$ of all occurrences of $P$ in $\D$, the frequencies for each distinct document in $\mDA[s_p..e_p]$ are computed using a rank array on the document array, i.e., the number of occurrence of $P$ in document $T_i$ are $\rank_{i}(\mDA, s_e) - \rank_{i}(\mDA, s_p-1)$. Using a {\em wavelet tree}~\cite{DBLP:conf/soda/GrossiGV03} to represent the document array, given a pattern $P[1..m]$, Problem~\ref{problem:listing+freq} can be solved in $\Oh(t_{search}(m) + \ndoc\log t)$ time.%

Belazzougui {\em et al.}~\cite{DBLP:journals/jda/BelazzouguiNV13} build a {\em monotone minimum perfect hash function}~\cite{DBLP:conf/soda/BelazzouguiBPV09} on the document array. Combining Muthukrishnan's~\cite{muthukrishnan2002efficient} and Sadakane's~\cite{sadakane2007succinct} approaches, it is possible to find the leftmost and rightmost occurrence of the pattern $P$ in the $i$-th document. Using the constant time rank on the document array, Problem~\ref{problem:listing+freq} can be solved in $\Oh(t_{search}(m) + \ndoc)$ time. 

Gagie {\em et al.}~\cite{DBLP:journals/tcs/GagieNP12} propose a solution based on {\em wavelet trees}~\cite{DBLP:conf/soda/GrossiGV03}, that does not rely on Muthukrishnan's~\cite{muthukrishnan2002efficient} solution. The idea is to use a the {\em range quantile}~\cite{DBLP:conf/spire/GagiePT09} problem to find the $i$-th smallest value in the range $\mDA[s_p..e_p]$. Then, retrieve its frequency as the length of interval corresponding to $[s_p..e_p]$ in its leaf in the wavelet tree. With this approach Problem~\ref{problem:listing+freq} can be solved in $\Oh(t_{search}(m) + \ndoc\log t)$ time.

\section{The document listing with frequencies} 
We are now ready to describe our {\em document listing with frequencies} approaches. We propose three different solutions, which rearrange and adapt different concepts of previous work. The first solution is based on the solution for the {\em document listing} proposed in~\cite{cobas2019fast}. We grammar compress \DA{}, and for all non terminal nodes, we precompute and store the results of {\em document listing with frequencies} queries. The second solution combines Sadakane's approach~\cite{sadakane2007succinct} for the {\em term frequency} problem, with the grammar compressed document array. We enhance the grammar compressed document array with bitvectors in each non terminal, to locate the leftmost and rightmost occurrences of each document in the corresponding interval in the document array. The third solution combines Sadakane's approach~\cite{sadakane2007succinct} for the {\em term frequency} problem, with the $\mILCP$ array. In this case we use two copies of the $\mILCP$ array to locate the leftmost and rightmost occurrences of each document in the corresponding interval in the document array.

As a common step in all three approaches, given a collection\\ $D = \{T_1[1..n_1], \ldots, T_t[1..n_t]\}$, we build one \ri{} for the concatenation of the documents $\D$. Given the pattern $P[1..m]$, in order to find the frequencies  of the occurrences of the pattern in each document, we first find all occurrences of the pattern $P$ in the concatenation of all documents $\D$ using the \ri{} in $\Oh(m)$ time and $\Oh(r\log(n/r))$ bits. All occurrences of the pattern $P$ are identified as an interval in the suffix array of $\D$, i.e. $\mSA[s_p..e_p]$.

For the second and the third approach we also build an \ri{} for each document $T_i$, for $1 \leq i \leq t$. The \ri{} for $T_1,\ldots,T_t$ can be built in $\Oh(\sum_{i=1}^t n_i) = \Oh(n)$ time and occupying $\Oh(\sum_{i=1}^t r_i\log (n_i/r_i)) = \Oh(Rt\log(n/r_k))$ bits, where $R = \sum_{i=1}^t r_i$ and $k = $argmin$(r_1, \ldots, r_t)$.

\subsection{Precomputed document list with frequencies} \label{GCDA-PDL}
Following the ideas for the {\em document listing} problem proposed in~\cite{cobas2019fast}, we grammar compress \DA producing a binary and balanced grammar of $\nu$ non-terminals, that can be stored in $\Oh(r\log(n/r))$ bits~\cite{Travis18}. Let $\T$ be the parse tree of the document array $\DA[1..n]$, given a non terminal node $nt \in \T$ let $\DA[s_{nt}..e_{nt}]$ be its expansion. For all non terminal nodes $nt \in \T$, we precompute and store the list $D_{nt}$ of the distinct documents in $\DA[s_{nt}..e_{nt}]$ with their frequencies. The lists are stored in ascending order. 

\subsubsection{Query.}
Given the range $[s_p..e_p]$ of all occurrences of $P$, we find maximal nodes of the parse tree $\T$ that cover $\DA[s_p..e_p]$. Since the grammar is binary and balanced, the number of maximal non terminal nodes covering $\DA[s_p..e_p]$ is $\Oh(\log n)$. Those nodes can be found in $\Oh(\log n)$ time traversing the parse tree $\T$ from the root towards the interval $\DA[s_e..s_p]$. We use an atomic heap~\cite{FW94} to merge the $\Oh(\log n)$ lists and compute the frequencies of the documents, by inserting the head of each list in the heap; extracting the minimum and inserting the next element from the same list. While extracting the document, we compute the frequencies for each document. The atomic heap allows to insert end extract the minimum in constant amortized time, thus the total time to compute the output is $\Oh(\ndoc\log n)$ since each document can appear in each list.

Summarizing, we can answer to Problem~\ref{problem:listing+freq} in $\Oh(m + \ndoc\log n)$ time, using $\Oh(r\log(n/r) + t \times \nu)$ bits.

\subsection{Grammar-compressed document array with bitvectors} \label{GCDA}
Let $\T$ be the parse tree of the document array $\DA[1..n]$ with $\nu$ non-terminals. %
For each non terminal node $nt \in \T$ we store if the $i$-th document occurs in the expansion of $nt$ and, if so, whether the leftmost (resp. rightmost) occurrence is in the left child or in the right child of $nt$. Let $\ell$ and $r$ be the left child and right child of $nt$, respectively. The above information can be stored into two bitvectors $\mL_{nt}$ and $\mR_{nt}$ of length $t$, such that for all documents $i = 1, \ldots, t$, $\mL_{nt}[i] = 0$ if the leftmost occurrence of the $i$-th document is in $\ell$, and $1$ otherwise, and $\mR_{nt}[i] = 1 $ if the rightmost occurrence of the  $i$-th document is in $r$, and $0$ otherwise. 
Note that if $\mL_{nt}[i] > \mR_{nt}[i]$, then the $i$-th document does not occur in $nt$.

For the $i$-th document it holds that $\mL_{nt}[i] = \mL_{\ell}[i]\wedge\overline{\mR_{\ell}[i]}$ and $\mR_{nt}[i] = \overline{\mL_{r}[i]} \vee \mR_{r}[i]$
where $\overline{x}$ is $1 - x$. We store $\mL_{nt}$ and $\mR_{nt}$ in each non terminal node and we compute them in a bottom up fashion. Considering that non terminal nodes associated to the same non terminal symbol have the same subtree, we can compute the $\mL_{nt}$ and $\mR_{nt}$ bitvectors only once for each non terminal symbol. Thus, the whole running time of the algorithm is $\Oh((t/w) \times \nu)$ using bit parallelism on words of $w$ bits.

\subsubsection{Query.} 

Let $t_1, \ldots, t_k$ be the $ k = \Oh(\log n)$ maximal non terminals that cover the interval corresponding to $\mDA[s_p..e_p]$.
We build a binary tree $\T'$ having as leaves the nodes corresponding to $t_1, \ldots, t_k$. Each internal node stores a pair of bitvectors $L$ and $R$, computed using the rules described above. The height of $\T'$ is $\Oh(\log\log n)$.
To retrieve the leftmost and rightmost occurrences of each document, we start from the root of $\T'$, for each document present in the root, we descend the tree, using the information stored in the bitvectors, to find first the leftmost, and then the rightmost occurrence of the document.

We perform exactly two traversals of the tree for each document that occurs at least once in the interval, since the $\mL$ and $\mR$ bitvectors stores the information that a document does not appear in the interval of the node.
Using bit parallelism on words of size $w$, we can find the leftmost and rightmost occurrence of each document in $\Oh(\ndoc(t/w) (\log n + \log\log n))$ time.

Once we have computed the leftmost and rightmost occurrences $\ell_i$ and $r_i$ for each document $i$, we use random access to $\mSA$ of the \ri{} to find their corresponding suffix values $\mSA[\ell_i]$ and $\mSA[r_i]$ in the concatenation of the documents. We, then, find the corresponding suffix values in the document $T_i$, and, using random access to $\mISA$ we find the interval the leftmost and rightmost occurrence $\ell_i'$ and $r_i'$ in the suffix array of the document $T_i$. The size of this interval is the number of occurrences of the pattern $P$ in $T_i$, i.e. $r_i' - \ell_i' + 1$.

Keeping all together, we can answer queries to Problem~\ref{problem:listing+freq} in $\Oh(m + ((t/w)\log n + \log(n/r))\ndoc)$ time, using $\Oh(r\log (n/r) + Rt\log(n/r_k) + (t/w) \times \nu)$ bits.

\subsection{Double run-length encoded $\mILCP$} \label{ILCP}

We first introduce a variation of the interleaved $\mLCP$ array introduced in~\cite{gagie2017document} called {\em double run-length encoded $\mILCP$}, denoted by $\mILCP^\bigstar$. The $\mILCP^\bigstar$ is composed by the array $\mVILCP^\bigstar$ storing the values of the runs, and the array $\mLILCP^\bigstar$ storing their lengths. Given the run-length encoded $\mILCP$ array for the collection $D=\{T_1,T_2. \ldots, T_t\}$ we merge together consecutive runs whose elements are from the same document, keeping the smallest value as the value of the run. Formally,  let $\rho$ e the number of runs of $\mILCP$, let $\ell_1 = 1$ and $r_1 = \mLILCP[1]$ for all $i=2,\ldots,\rho$ let $\ell_i = \sum_{j=1}^{i-1}\mLILCP[j]$ and $r_i = \ell_i + \mLILCP[i] -1$. Moreover, for all $1 \leq i \leq j \leq n$, let $|\DA[i..j]| = |\{\DA[k] \mid i \leq k \leq j\}|$ .

\begin{definition}
Let us assume that we have computed the run-length encoding up to position $i$ of $\mVILCP$, the next run of $\mILCP^\bigstar$ is defined as follows. Let $\ell =max\{k \mid |\mDA[\ell_i..r_k]| = 1\}$  if $|\mDA[l_i..r_i]| = 1$ and $0$ otherwise. Then $\mVILCP^\bigstar[j] = \min\{\mVILCP[i..i+\ell]\}$, and $\mLILCP^\bigstar = \sum_{k=i}^{i + \ell}\mLILCP[k]$.
\end{definition}

Extending~\cite[Lemma 1]{gagie2017document} to $\mILCP^\bigstar$ we have that: (proof in Appendix~\ref{app:proof lemma})
\begin{lemma}\label{lemma:ilcpast}
Given a collection $D = \{T_1, \ldots, T_t\}$ whose concatenation is $\D[1..n]$, let \SA be its suffix array, and let \DA be its document array. Let $\mSA[s_p..e_p]$ be the interval corresponding to the occurrences of the pattern $P[1..m]$ in $\D$. Then, the leftmost occurrences of the distinct documents identifiers in $\mDA[s_p..e_p]$ are in the same positions as the values strictly less than $m$ in $\mILCP^\bigstar[s_p..e_p]$. If there are two values smaller than $m$ for one document, we consider the leftmost one.
\end{lemma}

We build the double run-length encoded $\mLCP$ array on $\D$. We, then, build a {\em range minimum query} data structure~\cite{FH11} on $\mVILCP^\bigstar$ and a bitvector $\L[1..n]$ such that $\mLILCP^\bigstar[i] = \select_1(\mL,i)$. This allows, together with Lemma~\ref{lemma:ilcpast}, to use Sadakane's approach to find distinct documents to $\mVILCP^\bigstar$. This allows us to retrieve the leftmost occurrences of the distinct documents.
To retrieve the rightmost occurrence, we build the $\mILCP$ array using the {\em right} \LCP{}, i.e. the \LCP{} array defined as follows. We store in each position $1 \leq i \leq n-1$ the length of the longest common prefix between the two strings $T[\mSA[i]..n]$ and $T[\mSA[i+1]..n]$. In this case, we have that the rightmost occurrences of the distinct documents in $\DA[s_p..e_p]$ correspond to values of the $\mILCP$ strictly smaller than $m$. In particular, all properties that applies to the $\mILCP$ applies to the $\mILCP$ defined array using the {\em right} \LCP{}. We, then, also double run-length encode it.

\subsubsection{Query.}
Given the interval $[s_p..e_p]$, as in~\cite{gagie2017document}, we apply Sadakane's technique to find distinct elements in $\mDA$, to find distinct values in both the {\em double run-length encoded} $\mILCP$ arrays. Provided the positions of the leftmost and rightmost occurrences of each document, we then use the \ri{} to find the corresponding value of the suffix array. We map those positions back in the original document, and, using random access to $\mISA$ of the document, we obtain the interval $[s'_p..e'_p]$ in the suffix array of the document, whose size corresponds to the frequency of the document.

Keeping all together, we can answer queries to Problem~\ref{problem:listing+freq} in $\Oh(m + \log(n/r) \times \ndoc)$ time, using $\Oh(r\log (n/r) + Rt\log(n/r_k) + |\mILCP^\bigstar s|)$ bits, where $|\mILCP^\bigstar s|$ is the size of both the $\ILCP^\bigstar$ arrays.

\section{Experimental result}

We implemented the data structures and measured their performance on real-world datasets. Experiments were performed on a server with Intel(R) Xeon(R) CPU E5-2407 processors @ $\SI{2.40}{\GHz}$ and $\SI{250}{\gibi\byte}$ RAM running Debian Linux kernel \texttt{4.9.0-11-amd64}. The compiler was {\tt g++} version 6.3.0 with {\tt -O3} {\tt -DNDEBUG} options. Runtimes were recorded with Google Benchmark framework\footnote{github.com/google/benchmark}. The source code is available online at: \url{github.com/duscob/dret}

\subsubsection{Datasets.}
To evaluate our proposals, we experimented on different real and synthetic datasets. We used a variation of the dataset described by M{\"{a}}klin {\em et al.}~\cite{maklin2020genomic}, and some of the datasets tested by Cobas and Navarro~\cite{cobas2019fast}. These are available at {\tt zenodo.org} and {\tt jltsiren.kapsi.fi/RLCSA}, respectively. Table \ref{tab:collections} in Appendix~ \ref{app:tables} summarizes some statistics on the collections and patterns used in the queries.

\paragraph{Real datasets.} We used two repetitive datasets from real-life scenarios: \texttt{Species} and \texttt{Page}.
\texttt{Species} collection is composed of sequences of  {\em Enterococcus faecalis}, {\em Escherichia coli} and {\em Staphylococcus aureus} species. We created three documents, one per species, containing sequences of different strains of the corresponding species. We created two variants of \texttt{Species} dataset with 10 and 60 strains per document.
\texttt{Page} is a collection composed of pages extracted from Finnish-language Wikipedia. Each document groups an article and all its previous revisions. We tested on two variants of \texttt{Page} collection of different sizes: the smaller composed of 60 pages and  8834 revisions, and the bigger with 190 pages and 31208 revision.

\paragraph{Synthetic datasets.} Synthetic collections allow us to explore the performance of our solutions on different repetitive scenarios.
We experimented on the \texttt{Concat} datasets, very similar to \texttt{Page}. Each \texttt{Concat} collection contains $d = \lbrace10, 100\rbrace$ documents. Each document groups a base document and $10000/d$ versions of this. We generate the different versions of a base document with a mutation probability $R$. Notice that we have a \texttt{Concat} dataset for each combination of $d = \lbrace10, 100\rbrace$ and $R = \lbrace0.001, 0.003, 0.01, 0.03\rbrace$. A mutation is a substitution by a different random symbol. The base documents sequences of $\num{1000}$ symbols randomly extracted from English file of Pizza\&Chili~\cite{pizzachili}.

\paragraph{Queries.} The query patterns for \texttt{Species} collections are substrings of lengths $\vsPat = \lbrace 8, 12, 16 \rbrace$ extracted from the dataset. In the case of \texttt{Page} datasets, the patterns are Finnish words of length $\vsPat \ge 5$ that appears in the collections. For \texttt{Concat} collections, the queries are terms selected from an MSN query log. See Gagie \emph{et al.}~\cite{gagie2017document} for more details.

\subsubsection{Implementation details.}

All our implementations use the \ri{} as text index. We use the implementation of~\cite{Travis18} available at \url{github.com/nicolaprezza/r-index}. Since the implementation does not support random access to $\mSA$ and $\mISA$, we used a grammar-compressed \emph{differential suffix array} and \emph{differential inverse suffix array} --- the differential versions stores the difference between two consecutive values of the array ---. M{\"{a}}kinen {\em et al.}~\cite{MakinenNSV10:SRH} shows that SA of repetitive collections contains large \emph{self-repetitions} wich are suitable to be compressed using a grammar compressor like balanced Re-Pair.

Since we use the random access to $\mSA$ and $\mISA$ to retrieve the frequencies of the distinct documents, we implemented also a variant using a {\em wavelet tree} on the document array, as in~\cite{DBLP:conf/cpm/ValimakiM07}, to support the $\rank$ functionalities over $\mDA$.
For our experiments, we use the \texttt{sdsl-lite}~\cite{gbmp2014sea} implementation of the wavelet tree.

\subsubsection{Algorithms.}

We plugged-in our proposal with two different approaches to calculate the frequencies from the occurrences. 
All implementations marked with \idxName{-ISA} uses the random access to $\mSA$ and $\mISA$ to retrieve the frequencies, while the one marked with \idxName{-WT} uses the {\em wavelet tree}.

\begin{itemize}
    \item \idxName{GCDA-PDL}: {\em Grammar-Compressed Document Array with Precomputed Document Lists}. Solution described in Section~\ref{GCDA-PDL}, using balanced Re-Pair\footnote{www.dcc.uchile.cl/gnavarro/software/repair.tgz} for $\mDA$ and sampling the sparse tree as in~\cite{cobas2019fast}.
    
    \item \idxName{GCDA}: {\em Grammar-Compressed Document Array}. Solution described in Section~\ref{GCDA}, using balanced Re-Pair for $\mDA$ and bit-vectors stored in the non-terminals. We implemented the variants: \idxName{GCDA-ISAs} and \idxName{GCDA-WT}.
    
    \item \idxName{ILCP}: {\em Interleaved Longest Common Prefix}. Solution described in Section~\ref{ILCP}, using $\mILCP$ array (not double run-length encoded). We implemented the variants: \idxName{ILCP-ISAs} and \idxName{ILCP-WT}.
    
    \item \idxName{ILCP$^\bigstar$}: {\em double run-length encoded Interleaved Longest Common Prefix}. Solution described in Section~\ref{ILCP}, using $\mILCP^\bigstar$ array.  We implemented the variants: \idxName{ILCP$^\bigstar$-ISAs} and \idxName{ILCP$^\bigstar$-WT}.
    
    \item \idxName{Sada}: {\em Sadakane}. The algorithm proposed in~\cite{sadakane2007succinct}.  We provided the variants: \idxName{Sada-ISAs} and \idxName{Sada-WT}.
    
    \item \idxName{R-Index}: {\em \ri{}}. bruteforce algorithm that scans all occurrences of the pattern, counting the frequencies.
\end{itemize}

Note that in all our algorithms we do not use the random access to $\mSA$ and $\mISA$ of the \ri{}, thus we do not need to store the samples. The only exception is \idxName{R-Index} which needs the samples to compute the frequencies.

\subsubsection{Results.}

\begin{figure*}[t!]
	\centering
	\begin{subfigure}[b]{0.49\linewidth}
		\includegraphics[width=\textwidth]{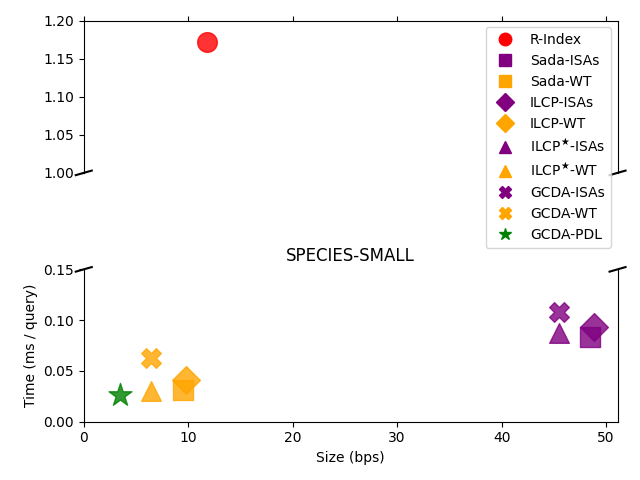}
	\end{subfigure}
	\begin{subfigure}[b]{0.49\linewidth}
		\includegraphics[width=\textwidth]{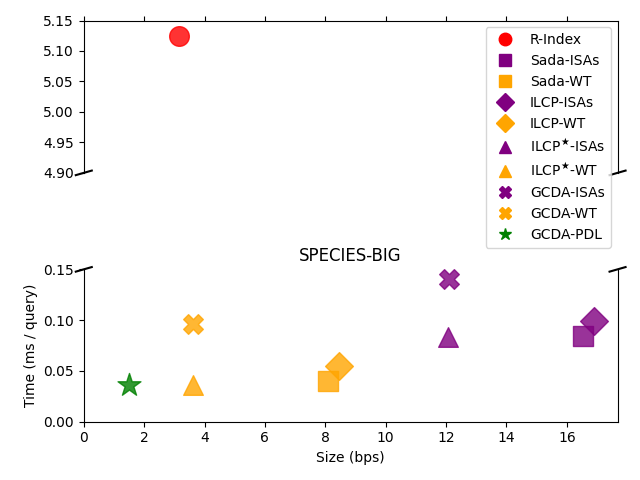}
	\end{subfigure}
	
	\centering
	\begin{subfigure}[b]{0.49\linewidth}
		\includegraphics[width=\textwidth]{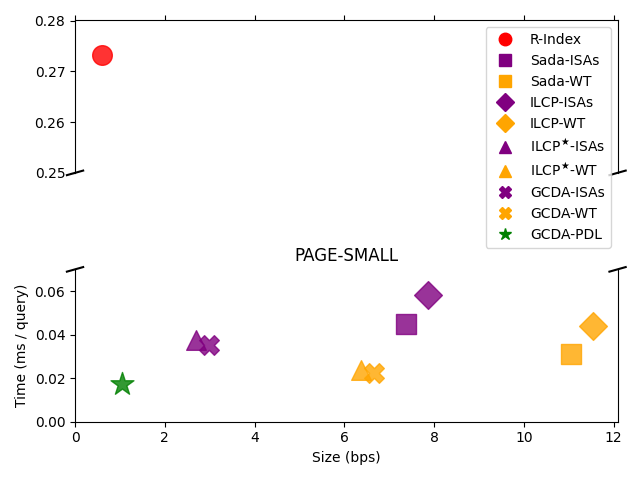}
	\end{subfigure}
	\begin{subfigure}[b]{0.49\linewidth}
		\includegraphics[width=\textwidth]{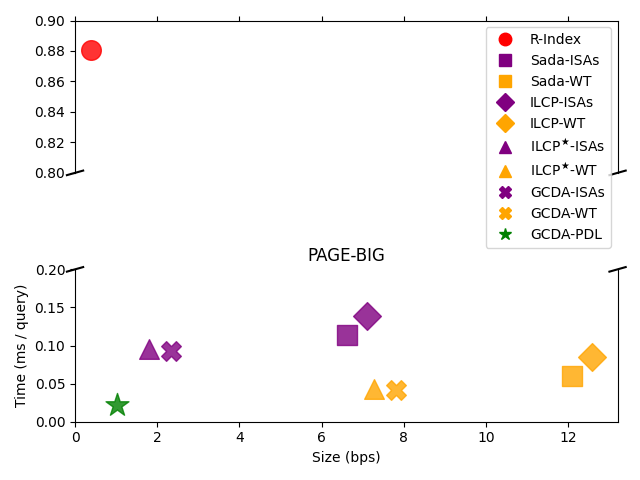}
	\end{subfigure}

	\caption{\small Document listing with frequencies on \texttt{Species} and  \texttt{Page} datasets. The $x$ axis shows the total size of the index in bps. The broken $y$ axis shows the average time per query.}
	\label{fig:real-coll}
\end{figure*}

Figure \ref{fig:real-coll} contains our experimental results for document listing with frequencies on real datasets. We show the trade-off between time and space for all tested indexes on different variants of the collections \texttt{Species} and \texttt{Page}.

The two variants of \texttt{Species} collections are composed of few large documents (only three, one per species). In this scenario, \idxName{GCDA-PDL} proves to be the best solution, finding the document frequencies in $\num{27}\text{--}\num{36}$ microseconds ($\mu$sec) per each pattern in average, and requiring only $\num{1.5}\text{--}\num{3.5}$ bits per symbols (bps). 
\idxName{GCDA-PDL} is the fastest and smallest index, requiring even less space than \idxName{R-Index}, since \idxName{GCDA-PDL} does not store the samples. The large size of the sampling scheme for collections with low repetitiveness has also been observed in~\cite{DBLP:journals/jacm/GagieNP20}.
The best competitor is \idxName{ILCP$^\bigstar$-WT}, being almost as fast ($\num{30}\text{--}\num{36}$ $\mu$sec per query) as \idxName{GCDA-PDL}, but requiring $\num{1.85}\text{--}\num{2.4}$ times more space. In these collections, \idxName{-WT} indexes perform better than \idxName{-ISAs} solutions. They can answer the queries at least $\num{1.45}$ times faster, while they are $\num{2}\text{--}\num{7}$ times smaller. In terms of space, \idxName{GCDA-WT} represents a good option, improving even the space required by \idxName{R-Index} in some cases, but much slower than \idxName{GCDA-PDL} and \idxName{ILCP$^\bigstar$-WT}.

\texttt{Page} collections that contain more documents than \texttt{Species} collections: $60$ documents in its small version and $190$ in the bigger one. Again \idxName{GCDA-PDL} turns up as the best index. It uses less than $\num{1.05}$ bps and answers the queries in $\num{17}\text{--}\num{22}$ $\mu$sec. \idxName{R-Index} requires the least space among the solutions, $\num{0.38}\text{--}\num{0.60}$ bps, but is $\num{15.86}\text{--}\num{40.35}$ times slower. The second overall-best index is \idxName{ILCP$^\bigstar$-ISAs}, with $\num{1.80}\text{--}\num{2.69}$ bps and query times of $\num{37}\text{--}\num{95}$ $\mu$sec, closely followed by \idxName{GCDA-ISAs}. On the \texttt{Page} variants, \idxName{-WT} indexes are faster than its counterparts \idxName{-ISAs}, but $\num{1.47}\text{--}\num{4.05}$ times bigger.

\begin{figure*}[t!]
	\centering
	\begin{subfigure}[b]{0.49\linewidth}
		\includegraphics[width=\textwidth]{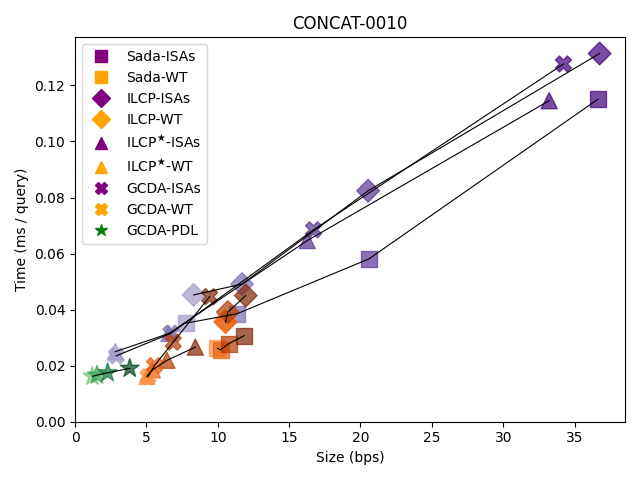}
	\end{subfigure}
	\begin{subfigure}[b]{0.49\linewidth}
		\includegraphics[width=\textwidth]{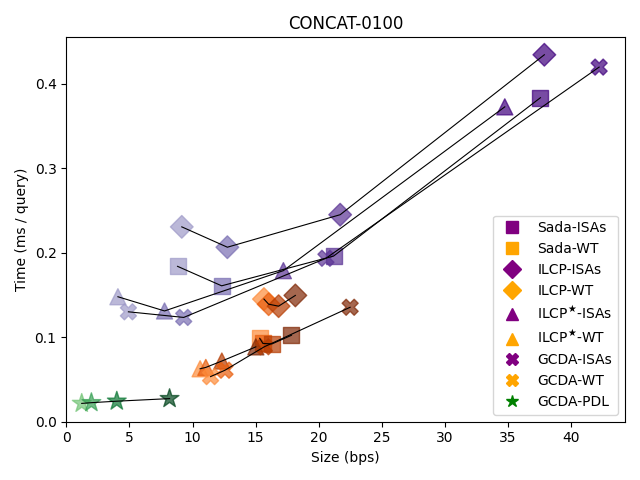}
	\end{subfigure}

	\caption{\small Document listing with frequencies on synthetic collection \texttt{Concat}. The $x$ axis shows the total size of the index in bps. The $y$ axis shows the average time per query. Indices with excessively high time are omitted in some plots.}

	\label{fig:coll_synthetic}
\end{figure*}

On real datasets \idxName{GCDA-PDL} outperforms the rest of the competitors, but the \idxName{ILCP$^\bigstar$-}variants are also relevant solutions obtaining a good space/time tradeoff.
The comparison of the indexes on synthetic collections \texttt{Concat} are shown in Figure \ref{fig:coll_synthetic}. These kinds of collections allow us to observe the indexes' behavior as the repetitiveness varies. Each plot combines the results for the different mutation probabilities of a given collection and number of base documents. The plots show the increasing mutation rates using variations of the same color, from lighter to darker.

\idxName{GCDA-PDL} outperforms all the other indexes. For the collections composed of $\num{10}$ base documents, our index obtains the best space/time tradeoff, requiring $\num{1.22}\text{--}\num{3.84}$ bps with a query time of $\num{16}\text{--}\num{19}$ $\mu$sec. Only \idxName{GCDA-WT} and \idxName{ILCP$^\bigstar$-WT} obtain competitive query times, but they are $\num{2.20}\text{--}\num{4.20}$ times bigger.
\idxName{R-Index} requires the least space for lower mutation rates, but it is $\num{79}\text{--}\num{83}$ times slower than \idxName{GCDA-PDL}. In the case of the collections composed of $100$ base documents, \idxName{GCDA-PDL} dominates the space/time map.

\section{Discussion\label{sect:discussion}}

Future work includes the integration of the results with real pseudoaligners. A trivial approach for such integration is to query each $k$-mer of a pattern with our methods, and check if a single document (species) receives positive term frequency. This approach multiplies the $O(m)$ part of the running time with $O(k)$, in addition to affecting the output-sensitive part of the running time. To avoid the $O(k)$ multiplier, we need to maintain the frequencies in a sliding window of length $k$ through the pattern. Such solution requires the techniques of the fully-functional bidirectional BWT index~\cite{BC19} extended to work on the $r$-index. However, one could also modify the pseudoalignment criterion into looking at maximal runs of $k$-mer hits, in the order of the (reverse) pattern. For this, our methods are readily applicable: Just do backward search with the pattern $P$ until obtaining an empty interval with suffix $P[i..m]$. Report term frequency of $P[i+1..m]$ if $m-i\geq k$. Continue analogous process  backward searching $P[1..i]$. If all the maximal runs of $k$-mer hits report a single document (species) $T_i$, assign $P$ to $T_i$. The $O(m)$ part of the running time remains unaffected, and the output-sensitive part remains smaller than with the sliding window approach.    

\bibliographystyle{splncs04}
\bibliography{bibliography}

\appendix
\section{Muthukrishnan's approach}\label{app:Muthukrishnan}

Muthukrishnan~\cite{muthukrishnan2002efficient} proposed the first solution to the document listing problem in optimal time and linear space. Given a collection $D$, the solution uses a suffix tree on the concatenation of all documents $\D$; the document array $\mDA[1..n]$; an array $\mC[1..n]$ which stores in each position $i$, the position in the suffix array of the suffix preceding $\mSA[i]$ in document $\mDA[i]$, i.e. $\mC[i] = j$ where $j<i$ is the largest position such that $\mDA[i]=\mDA[j]$, if such $j$ exists, we set $j=0$ otherwise; and a {\em range minimum query} data structure over $\mC[1..n]$ reporting, for each interval, the position in $\mC$ where the minimum occurs. Given the pattern $P$ of length $m$ we find the interval $\mSA[s_p..e_p]$ of all occurrences of $P$ in $\D$, using the suffix tree. All positions $i \in [s_p..e_p]$ such that $\mC[i] < s_p$ corresponds to distinct documents $\mDA[i]$. We find these positions using a recursive algorithm that, given an interval $[s_p.. e_p]$ first finds the position $i \in [s_p.. e_p]$ such that $\mC[i]$ is the minimum in $\mC[s_p.. e_p]$. If $\mC[i]\geq s_p$ then stop. Otherwise, reports $\mDA[i]$ and solve the same problem on the intervals $[s_p..j-1]$ and $[j+1..e_p]$ --- we always use the original value $s_p$ for the stop condition $\mC[i]\geq s_p$ ---. %

\section{Sadakane's approach}\label{app:Sadakane}

Sadakane~\cite{sadakane2007succinct} replaced the suffix tree with a compressed suffix array; the document array has been replaced by a bitvector $\mB[1..n]$ storing the position of the beginning of each document in text order, i.e. $\mB[i] = 1$ if $\D[i]$ is the first character of a document. Using a rank data structure over $\mB[1..n]$, then $\mDA[i] = rank_1(\mB,\mSA[i])$; the range minimum query over $\mC$ has been replaced by a succinct variant using $4n + o(n) $ bits, reduced to $2n + o(n)$ bits in~\cite{FH11}; the $\mC$ array has been removed and a bitvector marking the reported documents is used as stop condition of the recursive algorithm. %

\section{Gagie {\em et al.}'s approach}\label{app:Gagie et al}

Gagie {\em et al.}~\cite{gagie2017document} introduced the \ILCP array whose property stated in Lemma~\ref{lemma:ilcp} (Appendix~\ref{app:proof lemma}) allows to apply almost verbatim the technique used by Sadakane to find distinct elements in $\mDA[s_p..e_p]$. The solution uses a run-length compressed suffix array \RLCSA~\cite{MNSV10}; a bitvector $\mB[1..n]$ storing the position of the beginning of each document in text order; a bitvector $\mL[1..n]$ used as \LILCP, i.e. $\mLILCP[i] = \select_1(\mL,i)$; a  succinct range minimum query over \VILCP using $2\rho + o(\rho) $ bits; In order to solve the document listing problem we proceed as follows. Let $\mSA[s_p..e_p]$ be the interval of all occurrences of the pattern $P$ in $\D$, located using \RLCSA in $\Oh(t_{search}(m))$. We map the endpoints of this interval into the corresponding runs of the run-length encoded \ILCP, that are, $\ell = \rank_1(\mL,s_p)$ and $r = \rank_1(\mL,e_p)$. Apply Sadakane's technique to find distinct elements in $\mDA[s_p..e_p]$, to $\mVILCP[\ell..r]$. Each time we find a minimum in $\mVILCP[\ell..r]$, say in position $i$, we map that run in the original $\mILCP[\ell'..r']$ interval, where $\ell' = max(s_p, \select_1(\mL,i))$ and $r' = min(e_p, \select_1(\mL,i+1)-1)$. Then, for each position $\ell'\leq k \leq r'$ we compute $\mDA[k]$ using the bitvector $\mB$ and report it, marking the reported document bitvector. We iterate until we see a document that has already been marked.

\section{Proof of Lemma~\ref{lemma:ilcpast}}\label{app:proof lemma}

Here we are going to recall a nice property of $\mILCP$.
\begin{lemma}({\cite[Lemma 1]{gagie2017document}})\label{lemma:ilcp}
Given a collection $D = \{T_1, \ldots, T_t\}$ whose concatenation is $\D[1..n]$, let \SA be its suffix array, and let \DA be its document array. Let $\mSA[s_p..e_p]$ be the interval corresponding to the occurrences of the pattern $P[1..m]$ in $\D$. Then, the leftmost occurrences of the distinct document identifiers in $\mDA[s_p..e_p]$ are in the same positions as the values strictly less than $m$ in $\mILCP[s_p..e_p]$.
\end{lemma}

We are now going to show that this property can be extended to $\mILCP^\bigstar$.

\begin{proof}

For the runs of $\mILCP^\bigstar$ that are also runs of $\mILCP$, the property of Lemma~\ref{lemma:ilcp} holds.
We have to show that the same property holds also for runs of values from the same document.

Let $[s_p..e_p]$ be the interval of all occurrences of $P$ in the text. If a {\em same-document} run has value greater than or equals to $m$, then all occurrences in the run have $\ILCP$ value larger than or equals to $m$, hence by Lemma~\ref{lemma:ilcp} the property is satisfied. If the considered run has value strictly smaller than $m$ we have to consider three cases. The first case to consider is if the run is entirely included in $\mILCP[s_p..e_p]$, than the head of the run is the value strictly less than $m$, otherwise the head of the run would not be in the interval $\mILCP[s_p..e_p]$. The second case to consider is if the run is not entirely included in $\mILCP[s_p..e_p]$, and the run is broken by the left boundary of the interval, then, the leftmost occurrence of the document is in $s_p$. The last case is if the run is broken by the right boundary of the interval, then, if there is another run containing a value smaller than $m$ for document $i$, by Lemma~\ref{lemma:ilcp} the leftmost occurrence is the head of the other run, otherwise the leftmost occurrence is the head of the run crossing the right boundary. 
\end{proof}

Thus, considering the last run in the interval as a special case, we can apply the same approach as in~\cite{gagie2017document}. Then we consider the last run, checking if it is a {\em same-document} run or not, and if it is, we check if the same document has already been found by the algorithm.

\section{Missing tables\label{app:tables}}

\begin{SCtable}[][ht]
	\centering
	\small
	\begin{tabu}{l r r r r r}
		\toprule
		\rowfont{\em} Collection			& Size			& R-Index	& Docs			&Seqs		& Patterns \\

		\midrule
		\multirow{2}{*}{\texttt{Species}}	& 105			& 11.79		& \num{3}		& \num{10}		& \num{7658} \\
											& 631			& 3.15		& \num{3}		& \num{60}		& \num{20536} \\
		
		\cline{2-6}
		\multirow{2}{*}{\texttt{Page}}		& 110			& 0.60		& \num{60}		& \num{147}		& \num{7658} \\
											& 641			& 0.38		& \num{190}		& \num{164}		& \num{14286} \\
		
		\midrule
		\multirow{2}{*}{\texttt{Concat}}	& 95			&			& \num{10}		& \num{1000}	& \num{7538}\text{--}\num{10832} \\
											& 95			&			& \num{100}		& \num{100}		& \num{10614}\text{--}\num{13165} \\

		\bottomrule
	\end{tabu}
	
	\caption{ \scriptsize
		Statistics for document collections (small, medium, and large variants):
		\emph{Collection} name;
		\emph{Size} in megabytes;
		\emph{R-Index} bits per symbol (bps);
		\emph{Docs}, number of documents;
		\emph{Seqs}, average number of sequences (or versions) per each document;
		number of \emph{Patterns};
		For the synthetic collections (second group), we sum-up variants that use 10 or 100 base documents with the different mutation probabilities.
	}
	
	\label{tab:collections}
\end{SCtable}

\end{document}